\documentstyle[seceq,epsf,wrapft]{ptptex}
\title{%        %You can use \\ for explicit line-break
Study of Spectral Statistics of Classically Integrable Systems
}
\author{%       %Use \sc for the family name
Marko Robnik\footnote{e-mail: robnik@uni-mb.si} and Gregor Veble\footnote{
e-mail: gregor.veble@uni-mb.si}}

\inst{%         %Affiliation, neglected when [addenda] or [errata]
Center for Applied Mathematics and Theoretical Physics, University of 
Maribor\\
Krekova 2, SI-2000 Maribor, Slovenia}

\abst{%       %this abstract is neglected when [addenda] or [errata]
In this work we present the results of a study of spectral statistics for
a classically integrable system, namely the rectangle billiard. We show 
that the spectral statistics are indeed Poissonian in the semiclassical 
limit for almost all such 
systems, the exceptions being the atypical rectangles with rational 
squared ratio of its sides, and of course the energy ranges larger than 
$L_{\rm max}=\hbar / T_0$, where $T_0$ is the period of the shortest periodic
orbit of the system, however $L_{\rm max} \to \infty$ when $E \to \infty$.
}

\begin{document}

\maketitle

\section{Introduction}

The quantal energy spectra of classically integrable and classically 
chaotic systems show a remarkable difference in their statistical 
properties. The former appear to be random, satisfying Poissonian 
statistics, while the latter behave as eigenvalues of random orthogonal 
(Gaussian orthogonal ensembles or GOE) or Hermitian matrices 
(Gaussian unitary ensembles, GUE), depending on the properties of the 
system with respect to antiunitary transformations like the time reversal
\cite{rf:Bohigas1984,rf:BerryRobnik1986,rf:Robnik1986}.

These observations have, however, not yet been proven. There exists
numerical evidence supporting the Poissonian model for the 
integrable case \cite{rf:BerryTabor1977}. 
However, there has also been some evidence of the discrepancy 
of the Poissonian model for the case of the rectangle billiard 
\cite{rf:Casati1985,rf:Seligman1986,rf:Feingold1985}.

In this work we study the spectral statistics of a few integrable systems, 
namely the rectangle, torus and circle billiards. Here we present only the
results for the rectangle billiard. For results on the torus and circle 
billiards and a more in-depth discussion of the rectangle billiard see
reference \cite{rf:RobnikVeble1998}.

\section{Definitions}

We deal with an ordered spectrum $\{E_1, E_2,..., E_i,...\}$ of a given 
system. 
In order to compare spectral statistical properties of different systems
it is convenient to eliminate the system dependent average density of 
states. This is done by a nonuniform mapping of the energy scale called
the {\em unfolding of spectra}. If we know the average number of states up
to a given energy $\overline{N}(E)$, which can be obtained in
the first approximation by the Thomas-Fermi rule, then the mapping
\begin{equation}
x_i=\overline{N}(E_i)
\end{equation}
gives the unfolded spectrum $\{x_i\}$ 
of the system whose mean level spacing 
is equal to $1$.

\begin{wrapfigure}{r}{6.6cm}
  \epsfxsize = 2in 
 \centerline{\epsfbox{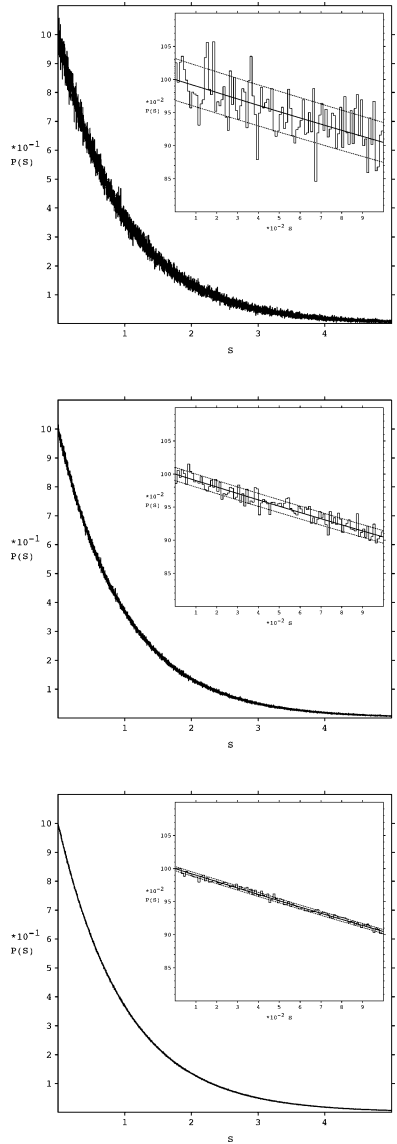}}
\caption{The level spacing distribution for different 
energy windows, top to bottom:
$10^6$ levels 
above unfolded energy $10^9$ (energy window $O_1$), 
$10^7$ levels above energy $10^{10}$ ($O_2$)
and $10^8$ levels above energy $10^{11}$ ($O_3$).
In the insert the appropriate 
magnifications for small level spacings are shown.}
\label{fig:1}
\end{wrapfigure}
The first measure of statistical spectral properties is the nearest 
neighbour level spacing distribution $P(S)$, which gives the distribution of
the distances between the consecutive energy levels. For Poissonian systems
it is (after unfolding)
\begin{equation}
P_{\rm Poisson}(S)=\exp(-S).
\end{equation}

The next measure of the spectral properties is the $\Delta$ 
statistics. It is
defined as
\begin{equation}
\Delta(L)=\frac{1}{L}\langle\min_{A,B} \int_{\alpha}^{\alpha+L} [N(x)-Ax 
-B]dx \rangle_{\alpha},
\end{equation}
where $N(x)$ is the spectral staircase function and counts the number 
levels up to energy $E$. This measure shows the average 
deviations of the spectral staircase function from the best fitting 
straight line over the interval $L$, and is therefore sometimes referred to as 
spectral rigidity. For systems with degeneracies of zero measure, for small 
$L$ it is always
\begin{equation}
\Delta(L)=L/15.
\end{equation}
For Poissonian spectra, however, this is true for all $L$.

Another spectral measure is important since all other measures can be 
derived from it. It is the $E(k,L)$ distributions, which give the 
probability of finding exactly $k$ levels in the interval of the length
$L$. Due to the unfolding of spectra, their maximum is close to $L\approx 
k$. For Poissonian systems,
\begin{equation}
E_{\rm Poisson}(k,L)=\frac{L^k}{k!} \exp(-L).
\end{equation}
The GOE/GUE expressions for the $E(k,L)$ statistics can be found in the 
form of tables in the reference \citen{rf:Mehta} for $k=0-7$, while for 
larger $k$ the asymptotic Gaussian formulae 
are applicable\cite{rf:Aurich1997}.

\section{Results}

The billiard we studied was the rectangle. Its spectrum is in appropriate
units given as
\begin{equation}
E_{m,n}=m^2+\alpha n^2,
\end{equation}
where $m,n$ are positive integers and $\alpha$ is the squared ratio of 
sides. 

\begin{wrapfigure}{r}{6.6cm}
\epsfxsize = 2 in
\centerline{\epsfbox{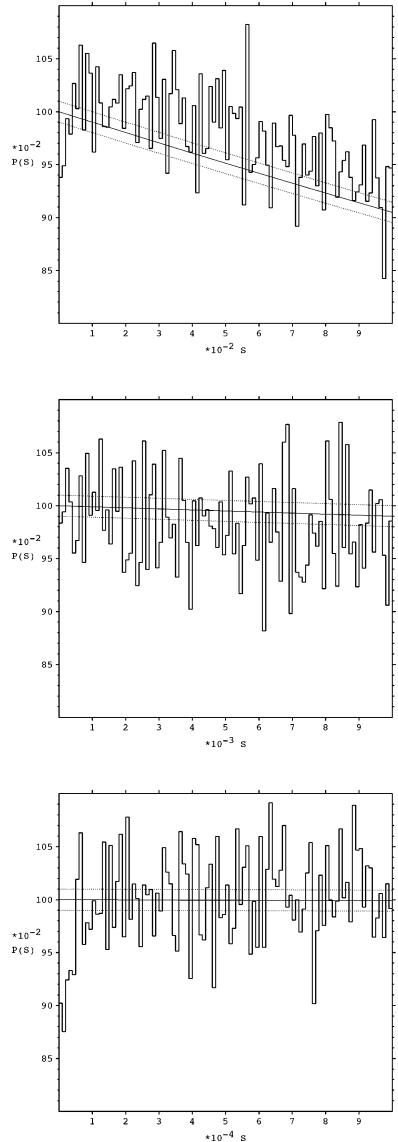}}
\caption{$P(S)$ for small level spacings for all levels up to energy (top
to bottom)
$10^7$, $10^8$ and $10^9$, 
with the bin size chosen such that
the expected number of events per bin is about $1000$.}
\label{fig:2}
\end{wrapfigure}
If $\alpha$ is rational,
\begin{equation}
\alpha=\frac{p}{q},
\end{equation} 
where $p$ and $q$ are positive integers, 
the energy levels become integer multiples of the quantity
\cite{rf:BerryTabor1977}
\begin{equation}
X_{p,q}=\frac{\pi}{4\sqrt{pq}}
\end{equation} 
in the units of the mean level spacing. This means that the level 
spacing distribution becomes a sum of delta functions separated by these 
steps. In reference \citen{rf:RobnikVeble1998} 
we show that even the amplitudes for the 
delta peaks do not follow the Poissonian prediction and that the degeneracy 
of levels increases with energy. 
Connors and Keating \cite{rf:Connors1997} proved that indeed the level
spacing distribution for the case of the square $\alpha=1$ billiard goes
towards
\begin{equation}
P_{\alpha=1}(S)=\delta(S),
\end{equation}
with $E\to\infty$.

Rational $\alpha$ are, however, not typical. For our studies we chose the
irrational 
$\alpha=\frac{\pi}{3}$, but also verified the results with the value of 
$\alpha=\frac{\sqrt{5}-1}{2}$, which is the golden mean and which gave 
qualitatively the same results.

In the figure \ref{fig:1} we plot the level spacing distribution for 
different high lying energy windows. We see that the Poissonian statistics
are followed excellently, with the statistical deviations remaining within
the expected values.

If, however, we try to include all the levels up to a given energy $E$ in 
the plot, and shrink the bin sizes so that at all energies there is the 
same expected number of events in the bin, we can see in the figure 
\ref{fig:2} that there exist
fluctuations around the Poissonian value which are larger than 
statistically expected and which do not seem to decrease with energy. 
We show in 
reference \citen{rf:RobnikVeble1998} 
that this happens due to the sufficiently influential 
rational $\alpha$ close to the chosen irrational value. If the bin size 
is kept constant, however, the distribution fluctuations tend to the 
statistically expected ones.
\begin{figure}[t]
\epsfxsize = 3 in
\centerline{\epsfbox{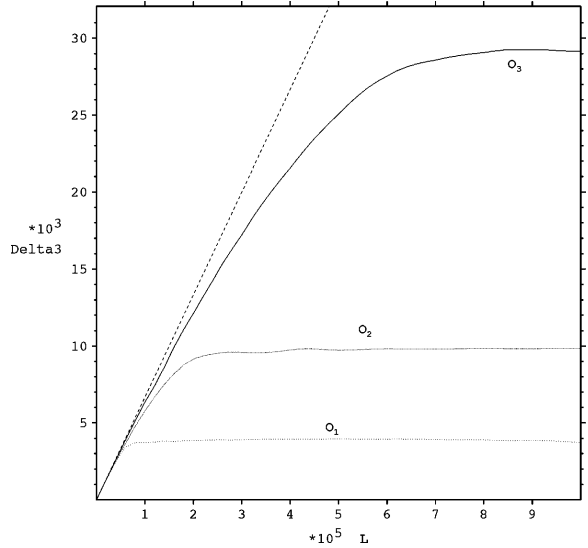}}
\caption{The $\Delta$ statistics for the same energy windows as in the
figure \protect\ref{fig:1}.}
\label{fig:3}
\end{figure}
\begin{figure}
\epsfxsize = 4 in
\centerline{\epsfbox{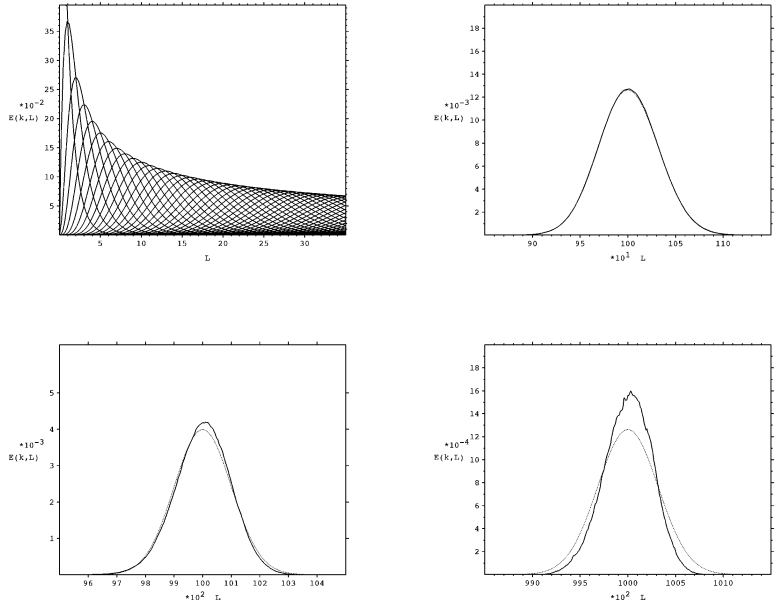}}
\caption{The $E(k,L)$ statistics for the energy
window $10^8$ levels above unfolded energy $10^{11}$ and $k=0..50$ (top left),
$k=10^3$ (top right), $k=10^4$ (bottom left) and $k=10^5$ (bottom right) }
\label{fig:4}
\end{figure}

The long range $\Delta$ statistics also behaves as expected, as can be 
seen in the figure \ref{fig:3}. For 
$L<L_{\rm max}$, where $L_{\rm max}=\hbar/T_0$ with $T_0$ being the
period of the shortest periodic orbit, $\Delta$ follows the 
Poissonian prediction. Berry \cite{rf:Berry1985} 
predicted through the use of the 
periodic orbit theory that at larger $L$ the $\Delta$ 
should saturate, as can also be confirmed in our plot. We may notice, 
however, that $L_{\rm max} \propto \sqrt{E}\to \infty$ as $E\to \infty$.

The $E(k,L)$ measures for different $k$ are plotted in figure \ref{fig:4}.
We can see that the Poissonian prediction is nicely followed up to 
$k$ being a fraction of the $L_{\rm max}$, 
where the numerical functions become narrower 
than the Poissonian ones. This can be understood by noting through the 
$\Delta$ plots that the spectrum becomes more rigid at higher $L$ than 
expected for the Poissonian one, so we should expect exactly $k$ levels 
to be found in a smaller ranges of $L$ around the average value 
$\langle L \rangle=k$. But, as $L_{\rm max}$ goes to infinity with 
increasing energy, the Poissonian $E(k,L)$ statistics are followed for 
all $k$ and $L$ in our system in the limit $E\to \infty$.

\section{Conclusion}

The Poissonian statistics are indeed nicely satisfied in the systems 
we studied,
apart from the atypical cases of the rational squared ratio of sides of the 
rectangle, and the energy ranges greater than $L_{\rm max}$. We show the 
same to be true also for the cases of the torus and circle billiards in 
reference \citen{rf:RobnikVeble1998}.

These studies are not important to understand the properties of 
the integrable systems only. In the case of the most general, mixed type 
systems, the total spectrum is in the semiclassical limit 
composed of independent contributions of the regular 
and chaotic components, with the regular component contribution 
having Poissonian and the
chaotic component contribution the GOE(GUE) statistics 
(the Berry-Robnik\cite{rf:BerryRobnik1984} theory), so a good understanding
of the individual components is necessary in order to understand the whole
picture. For a study of mixed type systems see reference \citen{rf:Veble1999}
and the short version in reference \citen{rf:Veble2000}

\end{document}